\documentclass[prl,
preprintnumbers,amsmath,amssymb]{revtex4}
\usepackage{graphicx}

\DeclareGraphicsExtensions{.pdf,*.jpg}

 \def\be{\begin{equation}}
 \def\ee{\end{equation}}
 
   \def\St{{\tilde{S}}}
  \def\Gt{{\tilde{G}}}
 \def\Rt{{\tilde{R}}}
 \def\lx{\lambda}

 \def\A{\mathcal{A}}

 \def\2{\frac{1}{2}}
 \def\4{\frac{1}{4}}

\catcode`\@=11
\def\@citex[#1]#2{%
\if@filesw \immediate \write \@auxout {\string \citation {#2}}\fi
\@tempcntb\m@ne \let\@h@ld\relax \def\@citea{}%
\@cite{%
  \@for \@citeb:=#2\do {%
    \@ifundefined {b@\@citeb}%
      {\@h@ld\@citea\@tempcntb\m@ne{\bf ?}%
      \@warning {Citation `\@citeb ' on page \thepage \space
undefined}}%
      {\@tempcnta\@tempcntb \advance\@tempcnta\@ne%
      \@tempcntb\number\csname b@\@citeb \endcsname \relax%
      \ifnum\@tempcnta=\@tempcntb 
it
        \ifx\@h@ld\relax%
          \edef \@h@ld{\@citea\csname b@\@citeb\endcsname}%
        \else%
          \edef\@h@ld{\ifmmode{-}\else--\fi\csname
b@\@citeb\endcsname}%
        \fi%
      \else
        \@h@ld\@citea\csname b@\@citeb \endcsname%
        \let\@h@ld\relax%
      \fi}%
    \def\@citea{,\penalty\@highpenalty\,}%
  }\@h@ld
}{#1}}

\def\@citeb#1#2{{[#1]\if@tempswa , #2\fi}}
\def\@citeu#1#2{{$^{#1}$\if@tempswa , #2\fi }}
\def\@citep#1#2{{#1\if@tempswa , #2\fi}}

\begin{document}

\title{Entanglement entropy, horizons and holography}

\author{D. Giataganas and N. Tetradis}
\affiliation{%
Department of Physics,
University of Athens,
University Campus,
Zographou, 157 84, Greece}
\date{\today}%

\begin{abstract}
We study the entanglement entropy in spaces with horizons, such as Rindler
or de Sitter space, using
holography. We employ appropriate 
parametrizations of AdS space in order to obtain a Rindler or static
de Sitter boundary metric. The holographic entanglement entropy for the regions
enclosed by the horizons can be identified with 
the standard
thermal entropy of these spaces. For this to hold, we define
the effective Newton's constant appropriately and account for the way the
AdS space is covered by the parametrizations.
\\
~\\
Keywords: Entropy; Holography.

\end{abstract}

\maketitle


We analyze the relation between 
entanglement entropy and the thermal entropy in spaces that contain horizons. 
We focus on the divergent part of the entanglement entropy, which is
known to scale with the area of the entangling surface \cite{sorkin}. This 
feature suggests a connection with the thermal entropy of the 
gravitational background when the 
entangling surface is identified with the horizon.
The framework for our analysis is provided by the 
AdS/CFT correspondence \cite{adscft}, 
which establishes a connection between an 
asymptotically anti-de Sitter (AdS) bulk space and a conformal field theory (CFT)
that can be considered as living on its boundary.
The metric on the AdS boundary belongs to a conformal class, a feature that
permits the study of the dual CFT on nontrivial backgrounds. We consider 
parametrizations for which the boundary metric takes the Rindler and static
de Sitter form.
We focus on the entanglement entropy associated with a CFT 
confined within a part $A$ of the 
AdS boundary delimited by an entangling surface $\A$. Our analysis is based on the proposal of refs. \cite{ryu,review} that the entropy is proportional to the
area of an appropriately defined minimal surface 
(holographic screen) $\gamma_A$ that starts from $\A$ and extends into the bulk. 

We consider three parametrizations of $(d+2)$-dimensional AdS space. 
The first uses global coordinates:
\be
ds^2_{d+2}
= \frac{R^2}{\cos^2\chi}\left[-d\tau^2+d\chi^2+\sin^2\chi  \left(d\theta^2+\sin^2\theta \, d\Omega^2_{d-1} \right)\right],
\label{global} \ee
with $-\infty< \tau < \infty$, $0\leq \chi < \pi/2$, $-\pi/2 \leq \theta \leq \pi/2$ 
for $d>1$, and $R$ the AdS radius. For $d=1$, $\theta$ covers the full unit
circle: $-3\pi/2\leq \theta \leq \pi/2$.
The second parametrization 
uses Fefferman-Graham coordinates \cite{fg} with a Rindler boundary:
\be
ds^2_{d+2}
= \frac{R^2}{z^2}\left[dz^2
-a^2 y^2 d\eta^2+dy^2 + d \vec{x}_{d-1} \right],
\label{rindler} \ee
with the boundary located at $z=0$, $a$ a constant parameter, 
$-\infty < \eta < \infty$,
$0< y<\infty$ covering the right (R) Rindler wedge and $-\infty < y <0$ the 
left (L) wedge. The third parametrization makes use of a 
de Sitter (dS) slicing of AdS space that results in a metric of the Fefferman-Graham
form with a static dS boundary:
\begin{equation}
ds^2_{d+2}
= \frac{R^2}{z^2} \left[ dz^2 +\left(1-\frac{1}{4}H^2 z^2 \right)^2 \left(
- (1-H^2\rho^2) dt^2 
+  \frac{d\rho^2}{1-H^2\rho^2}+\rho^2 \, d\Omega^2_{d-1} \right) \right],
\label{dS} \end{equation}
where $0\leq \rho \leq 1/H$ covers the static patch for $d>1$. There are two 
such patches in the global geometry, 
with $\rho=0$ corresponding to the ``North" and ``South poles".
For $d=1$, $\rho$ can be negative and each static
patch is covered by $-1/H \leq \rho \leq 1/H$. 
All the coordinates in the above expressions are taken to be dimensionless,
with $R$ the only dimensionful parameter. In particular, $H$ and $a$ are dimensionless. For de Sitter space, 
the physical Hubble scale is $H/R$.

There are several parametrizations
of AdS that correspond to different vacua of the boundary theory \cite{tetradis}.
The holographic stress-energy tensor of the dual CFT for the metric (\ref{rindler}), computed along the lines of \cite{skenderis}, vanishes. This indicates that
this form of the metric corresponds to the Minkowski vacuum, for which the
Rindler observer is expected to detect a thermal environment.
Similarly, the stress-energy tensor for the metric (\ref{dS}) does not contain
any singularities on the horizon. Therefore, this metric corresponds to the 
Bunch-Davies vacuum \cite{bunch} for which 
the static dS observer again detects a thermal 
background \cite{bd}. 

\begin{figure}[t]
\centering
$$
\includegraphics[width=0.42\textwidth]{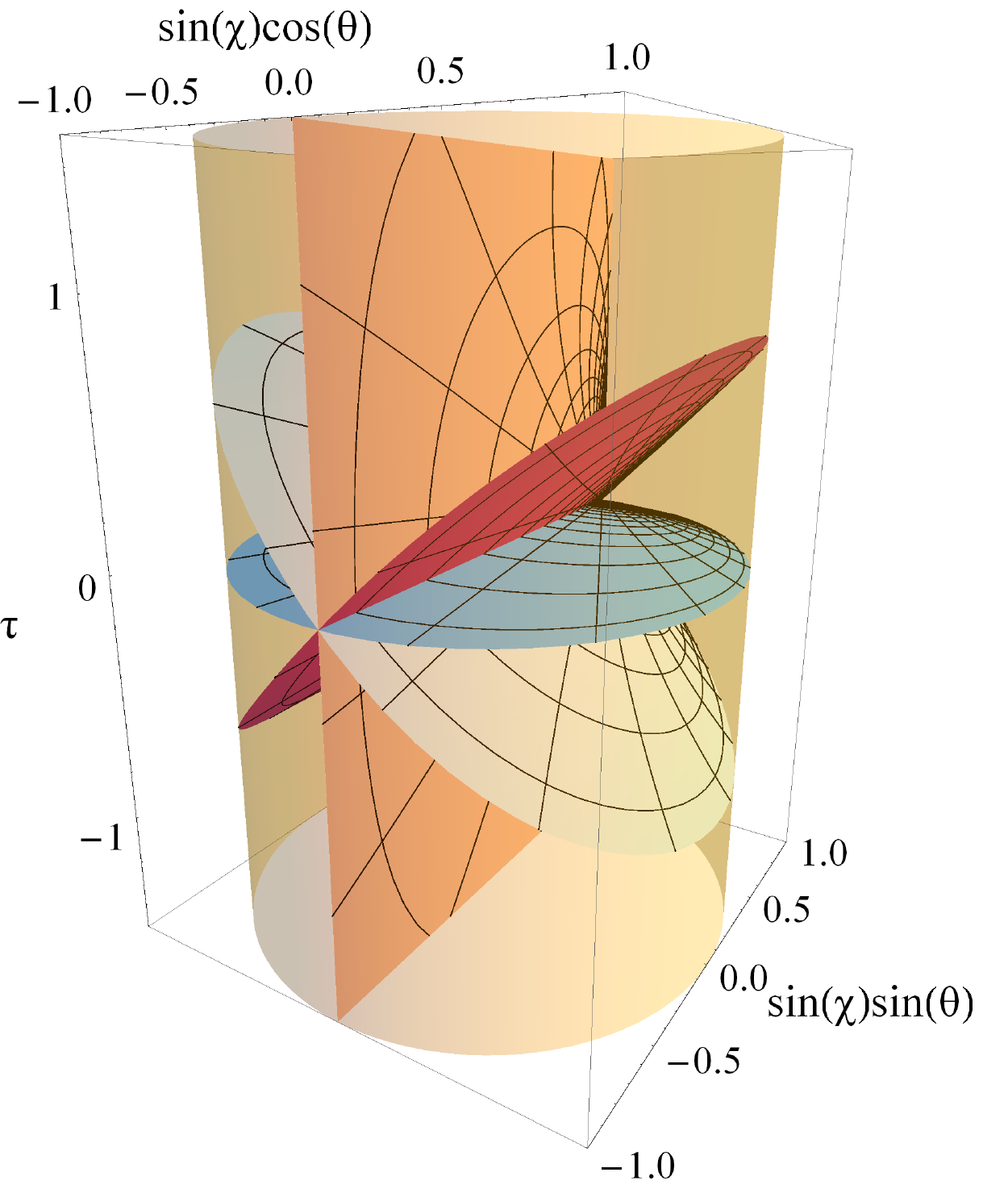}\hspace*{0.02\textwidth}
\includegraphics[width=0.42\textwidth]{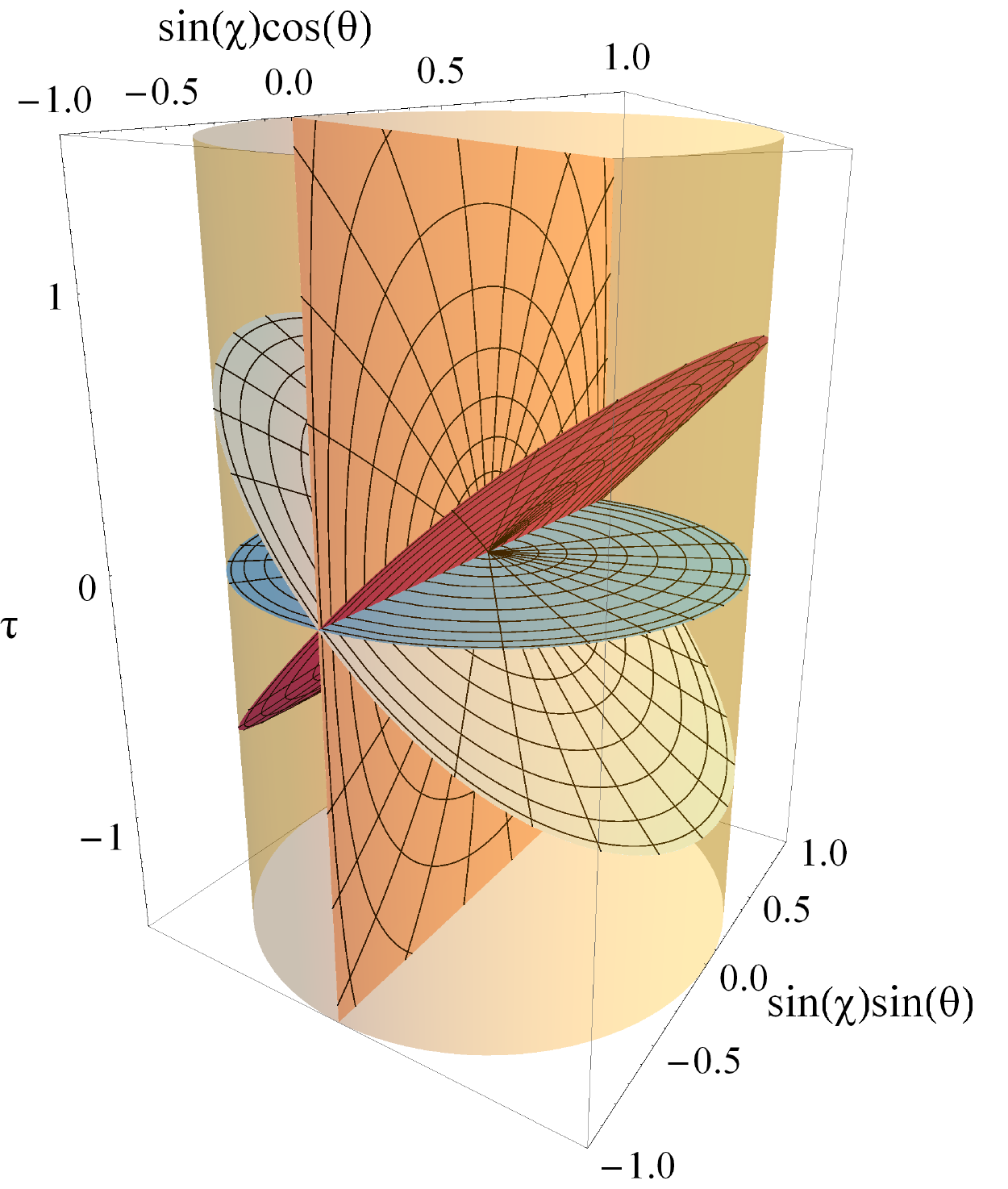}\hspace*{0.02\textwidth}
$$
\caption{The slicing of the Euclidean AdS$_3$ cylinder for a Rindler boundary with $a=1$
(left) and a static de Sitter boundary with $H=1$ (right). The Euclidean ``time" coordinates $\eta_E$ and $t_E$
are varied between $-\pi/2$ and $\pi/2$.}
\label{cylinder}
\end{figure}

We consider first the case of a Rindler boundary. 
The thermal nature of the vacuum is 
reflected in the periodicity of the Euclidean ``time" coordinate $\eta_E=i \eta$.
In order to understand how the parametrization covers the AdS space,
we consider the relation between the Euclidean Rindler coordinates 
and the Euclidean global ones. 
Similar conclusions can be reached by considering metrics with Lorentzian signature.
We use AdS$_3$ as an example, for 
which the explicit relation is given by 
\begin{eqnarray}
\chi(z,\eta_E,y)&=&\tan^{-1}\left(\frac{1}{z}
\sqrt{y^2 \cos^2(a\, \eta_E) +\frac{1}{4}\left(y^2+z^2-1 \right)^2}\right)
\label{yg} \\
\tau_E(z,\eta_E,y)&=&\tanh^{-1}\left(\frac{2y\sin(a\,\eta_E)}{y^2+z^2+1} \right)
\label{tg} \\
\theta(z,\eta_E,y)&=&\tan^{-1}\left(\frac{y^2+z^2-1}{2y\cos(a\,\eta_E)} \right).
\label{thg} \end{eqnarray}
The slicing of the Euclidean AdS$_3$ cylinder by these coordinates is depicted in
the left plot of 
fig. \ref{cylinder} for $a=1$. 
Each one of the depicted slices corresponds to a constant value
of $\eta_E$. The horizontal slice corresponds to $\eta_E=0$, the 
vertical one to $\eta_E=\pm \pi/2$ and the other two to $\eta_E=\pm \pi/4$. The 
intersection of each slice with the boundary corresponds to $z=0$. Each 
intersection is
parametrized by $y$, starting from the point $y=0$ at the front of the plot, which
corresponds to ($\chi=\pi/2$, $\theta=-\pi/2$) in global coordinates. One moves to 
the right or left on the boundary 
for positive or negative $y$, respectively. 
For $y\to + \infty$ one approaches the point 
($\chi=\pi/2$, $\theta = \pi/2$) at the back. For $y\to -\infty$ one approaches
the same point, now corresponding to ($\chi=\pi/2$, $\theta =-3 \pi/2$).
It is important for the following 
to keep in mind that ($\chi=\pi/2$, $\theta = \pi/2$) 
does not represent a point in Rindler space and that the limits 
$y\to \pm \infty$ must be considered distinct. 
Each slice of constant $\eta_E$ is covered by lines with $y$ constant and $z$
growing from $z=0$. For all values of $y$ and $z\to \infty$
these lines converge towards the point ($\chi=\pi/2$, $\theta = \pi/2$) on the boundary. The Rindler horizon corresponds to the boundary point 
($\chi=\pi/2$, $\theta = -\pi/2$) at the front. 
The intersection of all slices is the 
holographic image of the horizon that acts as a bulk horizon.
This construction maps the right Rindler wedge to the front half of the
AdS$_3$ cylinder in fig. \ref{cylinder}. The left Rindler wedge is mapped to the 
other half of the cylinder. With respect to entanglement entropy, 
the bulk horizon is also the minimal surface $\gamma_A$
of refs. \cite{ryu,review} for a boundary
region that includes the whole range $0<y<\infty$ for fixed $\eta_E$ and is 
entangled with  the region $0>y>-\infty$.

The calculation of the length of the minimal surface for a ($d+1$)-dimensional
Rindler boundary is identical to
the one in \cite{review} for an infinite belt on a Minkowski boundary. 
The surface $\eta=0$ corresponds
to Minkowski time $t=0$. On this surface, the coordinate $y=x>0$ covers the positive 
$x$-axis of Minkowski space, and $y=x<0$ the negative $x$-axis. The two 
regions are separated by the horizon at $y=0$.
We consider an infinite strip along the $y$-axis with width $l$. The minimal
surface extends into the bulk up to a turning point at 
$z_*=\Gamma\left(\frac{1}{2d}\right)/\left(2\sqrt{\pi}\,
\Gamma\left(\frac{d+1}{2d}\right)\right)\,l.$ The entanglement entropy is obtained by dividing the area of
the minimal surface by $4G_{d+2}$, with $G_{d+2}$ the Newton's constant of the
bulk space. The entanglement entropy is given by 
\begin{eqnarray}
S_A&=&\frac{2R(R^{d-1}L^{d-1})}{4G_{d+2}}
\left(
\frac{1}{(d-1) \epsilon^{d-1}}
-2^{d-1}\pi^{d/2}\left(\frac{\Gamma\left(\frac{d+1}{2d}\right)}{
\Gamma\left(\frac{1}{2d}\right)}\right)^d
\frac{1}{(d-1)l^{d-1}}
\right) 
\nonumber \\
&=&
\frac{2R(R^{d-1}L^{d-1})}{4G_{d+2}}
\left(
\frac{1}{(d-1) \epsilon^{d-1}}
+\frac{\sqrt{\pi}}{2d}\frac{\Gamma\left(\frac{1-d}{2d}\right)}{
\Gamma\left(\frac{1}{2d}\right)}
\frac{1}{z_*^{d-1}}
\right),
\label{Rindlerentropy} \end{eqnarray}
with $\epsilon$ a cutoff imposed on the bulk coordinate $z$ as the surface
approaches the boundary.
For $d=1$, one must substitute $1/((d-1)\epsilon^{d-1})$ with $\log(1/ \epsilon$), and
similarly for $l$ in the first line. The term in the parenthesis becomes 
$\log(l/\epsilon)$.
Here $L$ is the large length of the
directions perpendicular to the strip, so that $R^{d-1}L^{d-1}$ is the 
corresponding volume.

We are interested in the limit of eq. (\ref{Rindlerentropy}) when
the width of the strip covers the whole positive axis. The entropy
would arise from the entanglement of the right wedge with the left wedge of
Rindler space. As $z_*\to \infty$ for $l\to \infty$, it is apparent 
from the second line of eq. (\ref{Rindlerentropy}) that only the first 
term in the parenthesis survives
in this limit. This term is strongly dependent on the 
cutoff and independent of the width $l$ of the strip.
It must be noted, however, that for $d=1$ 
the first and second terms combine into $log(l/\epsilon)$.  
In the context of our considerations, it is clear that
the first term must 
have a physical meaning. Its nature becomes apparent if we define the effective
Newton's constant for the boundary theory following \cite{hawking}:
\be
G_{d+1}=(d-1) \epsilon^{d-1}\frac{G_{d+2}  }{R},
\label{Geff} \ee
with $(d-1)\epsilon^{d-1}$ replaced by 1/$\log(1/ \epsilon$) for $d=1$.
This definition is natural
within an effective theory that implements consistently a cutoff procedure
by eliminating the part of AdS space corresponding to $z<\epsilon$. Such
a framework is provided, for example, by the Randall-Sundrum (RS) model \cite{rs}.
If the positive-tension brane in this model is placed at $z=\epsilon$,
the massless graviton is localized arbitrarily close to the boundary, so
that the effective Newton's constant has the parametric dependence of
eq. (\ref{Geff}). As pointed out in \cite{hawking}, 
the cutoff dependence of the four-dimensional Newton's constant is not apparent 
in \cite{rs} because the metric is rescaled by $\epsilon^2$.
Our definition differs by a factor of 2 from the one in \cite{hawking}, in which 
the brane is assumed to bound two mirror AdS regions.
On the contrary, we integrate over the AdS region $z>\epsilon$ only once.
In a construction along the lines of the RS model, $G_{d+1}$ would be larger by a 
factor of 2, but the entropy would also be doubled because of the
presence of two copies of AdS space. 
Our definition does not include the factor of 2 and can be viewed
as resulting from the regulated action in the 
context of holographic renormalization \cite{skenderis,papadimitriou}.

The use of the definition (\ref{Geff}) in eq. (\ref{Rindlerentropy})
results in an entanglement
entropy $S_R=(R^{d-1}L^{d-1})/(2G_{d+1})$, which is double the
expected value of the thermal entropy \cite{laflamme}. 
This discrepancy can be understood by examining more carefully the limit
in which the strip covers the entire right wedge. The strip extends from 
the horizon at $y=0$ to a value $y_m$ for which the limit $y_m\to \infty$ is taken.
For any finite, no matter how large, value of $y_m$ the strip is
entangled not only with the left wedge, but also with the (infinite domain)
beyond $y_m$. As the space is essentially flat, the two contributions are
expected to be equal. If one is interested in the entanglement with
the left wedge only, the limit $y_m \to \infty$ must be accompagnied by a division
by 2 of the computed entanglement entropy.
The final result for
the Rindler entropy is 
\be
S_R=\frac{R^{d-1}L^{d-1}}{4G_{d+1}},
\label{entropyRindler} \ee
in agreement with \cite{laflamme}.

There is a natural identification of the entanglement entropy with the thermal 
one, resulting from the fact that 
the minimal surface $\gamma_A$ corresponding to the right wedge 
is nothing but the bulk horizon. 
In the left plot of fig. \ref{cylinder}, this is the line starting 
at the point ($\chi=\pi/2$, $\theta=-\pi/2$)
and extending towards the point ($\chi=\pi/2$, $\theta=\pi/2$).
As we pointed out earlier,
this second point does not correspond to one in Rindler space and is not 
a horizon of the boundary metric.
Therefore, the (infinite) contribution to the length of the
line from the vicinity of ($\chi=\pi/2$, $\theta=\pi/2$) should not be taken
into account. This interpretation justifies the division by
of 2 from a different point of view.

We turn next to the determination of the entropy of de Sitter space through
holography, a problem that has been the focus of several studies
\cite{hawking,dsprevious,myers,li,giataganas,nguyen,kiritsis,dSdS}.
Superconformal field theories in four-dimensional de Sitter space-time 
have been constructed in \cite{anous} and their properties 
have been studied in \cite{chu} through holography. 
We present here an explicit expression that 
interpolates between the entanglement entropy for regions much
smaller than the horizon and the thermal entropy of de Sitter space detected by
a static observer. The interpolation, which is possible through the
definition (\ref{Geff}) of the effective Newton's constant, makes more 
robust the relation between entanglement and thermal entropy of spaces with
horizons.

In order to make an explicit comparison with the Rindler case we 
consider how the Eudlidean AdS$_3$ is covered by the coordinates for a 
Euclidean de Sitter boundary.
The explicit relation is
\begin{eqnarray}
\chi(z,t_E,\rho)&=&\tan^{-1}\left( \frac{1-\frac{1}{4}H^2z^2}{H z}
\sqrt{\cos^2(H t_E) +H^2\rho^2 \sin^2(H t_E) }\right)
\label{ygds} \\
\tau_E(z,t_E,\rho)&=&\tanh^{-1}\left( \frac{1-\frac{1}{4}H^2z^2}{1+\frac{1}{4}H^2z^2} \sin(H t_E) \sqrt{1-H^2\rho^2} \right)
\label{tgds} \\
\theta(z,t_E,\rho)&=&\tan^{-1}
\left(\frac{H\rho}{\sqrt{1-H^2\rho^2}\, \cos(H t_E) } \right).
\label{thgds} \end{eqnarray}
In fig. \ref{cylinder} we depict the slicing of the AdS$_3$ cylinder for $H=1$.
Each one of the depicted slices corresponds to a constant value
of $t_E$. The horizontal slice corresponds to $t_E=0$, the 
vertical one, consisting of two parts, to $t_E=\pm \pi/2$, 
and the other two to $t_E=\pm \pi/4$. The 
intersection of each slice with the boundary corresponds to $z=0$. Each 
intersection has two symmetric regions, corresponding to the two
static patches of the global dS$_2$. Each one is
parametrized by $\rho$, starting from $\rho=-1/H$ at the front,
at the point ($\chi=\pi/2$, $\theta=-\pi/2$), and finishing at the
back, at the point ($\chi=\pi/2$, $\theta=\pi/2$) or 
($\chi=\pi/2$, $\theta=-3\pi/2$), for $\rho=1/H$. 
Each slice is covered by lines of constant $\rho$ and $z$.
These converge to the same point at the center, with $\chi=0$, 
for $z\to 2/H$. 
With respect to entanglement entropy, the minimal surface  $\gamma_A$
for each static wedge corresponds to the intersection of all slices, which connects
the two horizons at $z=0$, $\rho=\pm 1/H$. The entanglement occurs between the two
static de Sitter patches. The important difference with Rindler space is that
the endpoints of the minimal surface correspond to points of 
dS$_{2}$, they are actually the horizons of this space. So, no 
superfluous factor of 2 is expected. The situation is similar for $d>1$, with
the minimal surface  $\gamma_A$ 
ending on an $(d-1)$-dimensional sphere that separates the two
hemispheres of the $t_E=0$ slice of dS$_{d+1}$ \cite{kiritsis}. 

The equivalence between entanglement and thermal entropies is again apparent. 
The minimal surface is
nothing but the bulk horizon of the metric (\ref{dS}), extending between 
$z=0$ and $z=2/H$ for $\rho=\pm 1/H$, and between $\rho=-1/H$ and $\rho=1/H$
for $z=2/H$. This last part is actually one point, corresponding to
$\chi=0$ in global coordinates. For $d>1$ the situation is very similar, but now
$\rho>0$ and the cosmological dS horizon is the sphere with $\rho=1/H $. The
entanglement takes place between degrees of freedom on 
the two static patches. The holographic images of the patches are 
separated by 
the minimal surface starting from 
the dS horizon on the boundary \cite{hawking,kiritsis}.
This surface acts again as a bulk horizon.

For the calculation of the entanglement entropy for a general dimension $d$, 
we use the parametrization of eq. (\ref{dS}) and consider the interior $A$ of 
a spherical entangling surface $\cal A$ on the boundary. (For $d=1$ we consider a line segment between the two
horizons at $\rho=\pm 1/H$.) The minimal surface $\gamma_A$ in the bulk can be determined through the minimization of the area
\be
{\rm Area}(\gamma_A)=R^d S^{d-1}\int d\rho\, \rho^{d-1}
\frac{\left(1-\frac{1}{4}H^2z^2\right)^{d-1}}{z^d}
\sqrt{\frac{\left(1-\frac{1}{4}H^2z^2\right)^{2}}{1-H^2\rho^2}
+\left(\frac{dz(\rho)}{d\rho}\right)^2  },
\label{areads} \ee
with $S^{d-1}$ the volume of the ($d-1$)-dimensional unit sphere.
Through the definitions $\sigma=\sin^{-1}(H\rho)$, $w=2\tanh^{-1}(Hz/2)$, the 
above expression becomes
\be
{\rm Area}(\gamma_A)=R^d S^{d-1}\int 
d\sigma \frac{\sin^{d-1}(\sigma)}{\sinh^{d}(w)}\sqrt{1+\left(\frac{dw(\sigma)}{d\sigma}\right)^2}.
\label{areads2} \ee
Minimization of the area results in the differential equation
\be
\tan(\sigma)\tanh(w)\,w''+(d-1)\tanh(w)\left(\left(w'\right)^3+w'\right)
+d\tan(\sigma)\left(\left(w'\right)^2+1 \right)=0.
\label{diffds} \ee
For $H\to 0$ and small $\sigma$, $w$ we recover the known equation for 
the minimal surface in the case of a Minkowski boundary. Its solution is
$z(\rho)=\sqrt{\rho_0^2-\rho^2}$, with $\rho_0$ the radius of the
entangling surface on the boundary. 
The expressions for the area of the minimal surface and the corresponding
entanglement entropy are given in \cite{review}.

\begin{figure}[t]
\centering
\includegraphics[width=0.5\textwidth]{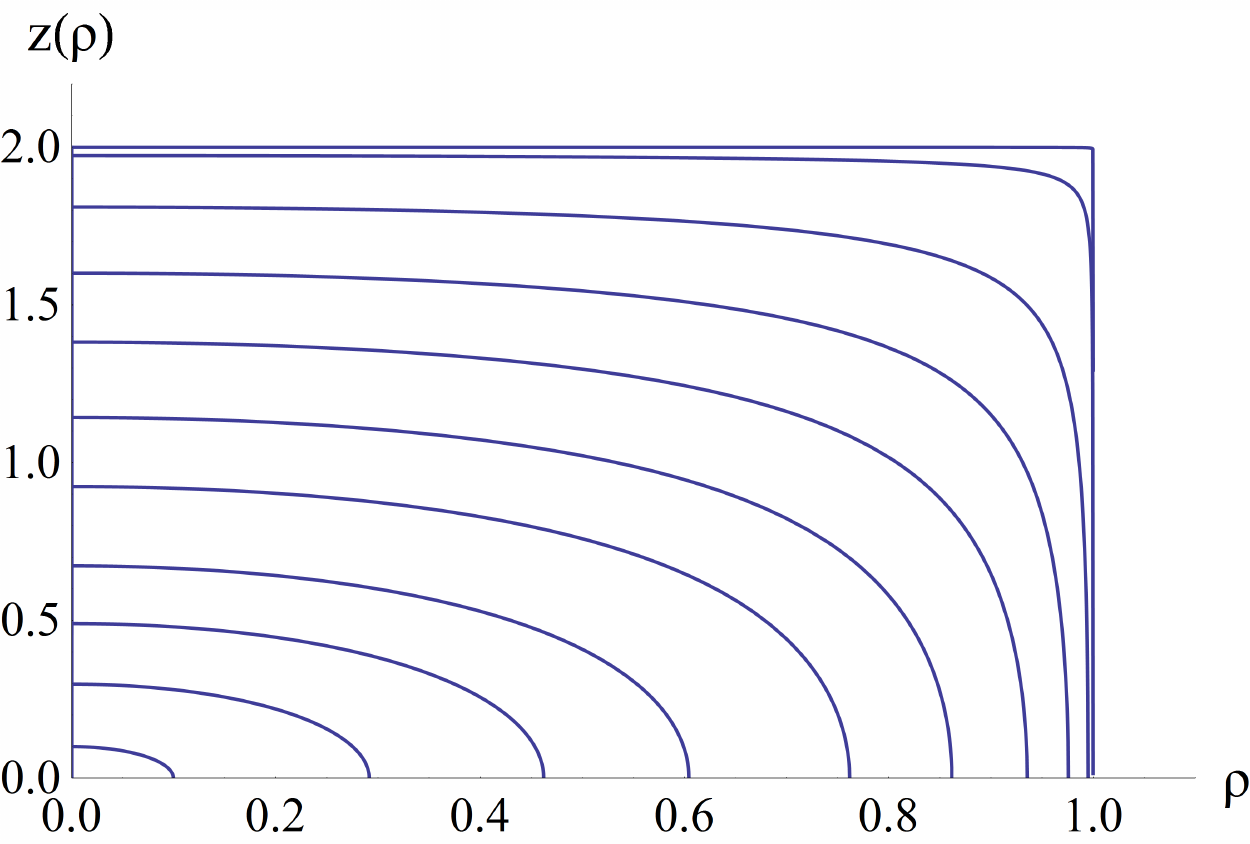}
\caption{Minimal surfaces for a de Sitter boundary with $H=1$.}
\label{solution}
\end{figure}

For nonzero $H$ and a general $d$, an analytical solution of 
eq. (\ref{diffds}) in closed form is not possible. However, the features
that are important for our analysis become visible in a numerical solution.
In fig. \ref{solution} we depict the solution for $d=3$, $H=1$ and increasing values of $\rho_0$. It is apparent that as the entangling surface approaches the horizon
the minimal surface has a limiting shape. Deep in the bulk it traces the 
surface with $z=2/H$, until the parameter $\rho$ reaches the value $\rho=1/H$.
Then it approaches the boundary with constant $\rho$ and diminishing $z$.
It is noteworthy that the first part of the surface has vanishing area, as
can be seen from eq. (\ref{areads}). The reason is that this part corresponds
to a single point in AdS space (the point with global coordinate $\chi=0$). The 
total area is dominated by the region near the boundary. For $\rho_0=1/H$,
the integration of eq. (\ref{areads}) is complicated by the singularity in $\rho$.
However, the expression (\ref{areads2}) demonstrates that the integral is
dominated by the region in which $w\simeq H z \to 0$, $dw/d\sigma \to -\infty$
and $\sigma \to \pi/2$. Cutting off the range of $z$ at $z=\epsilon$, gives
\be
{\rm Area}(\gamma_A)=R^d S^{d-1}\int_{H\epsilon} 
\frac{dw}{w^{d}}=\frac{R^d S^{d-1}}{(d-1)H^{d-1}\epsilon^{d-1}}.
\label{areaeps} \ee 
The entropy becomes 
\be
S_{dS}=\frac{R^d S^{d-1}}{4G_{d+2}(d-1)H^{d-1}\epsilon^{d-1}}=
\frac{S^{d-1}}{4G_{d+1}}\left(\frac{R}{H}\right)^{d-1},
\label{entropydss} \ee
which reproduces the thermal entropy of \cite{gibbons} 
and agrees with the findings of \cite{nguyen}.
The above expressions are valid for $d=1$ as well, 
with $1/((d-1)\epsilon^{d-1})$ replaced by $\log(1/ \epsilon$).

The general form of the expression for the thermal entropy raises
the question whether the identification with the entanglement 
entropy is also valid for bulk gravitational theories with
higher-derivative corrections. A typical example is 
Gauss-Bonnet (GB) gravity, which is dual to a CFT with a more
general class of central charges \cite{myers1}. The theory admits 
an AdS bulk solution of the form
\be
ds^2_{d+2}
= \frac{R^2}{z^2}\left[\frac{dz^2}{f}
-dt^2+d \vec{x}_{d} \right],
\label{GB} \ee
where $f=(1+\sqrt{1-4\lambda})/(2\lambda)$, with $\lambda$ the GB 
coupling in the notation of \cite{myers1}. The AdS radius is
equal to $\Rt=R/\sqrt{f}$. The RS construction for a GB bulk was carried
out in \cite{myers2}, with the brane located at $z=1$ in units of $\Rt$. 
Within our approach the brane is positioned at $z=\epsilon$ in units of
$\Rt$, which is equivalent to the rescaling of the induced metric
by a factor of $\epsilon^{-2}$. The induced Einstein term on the brane 
is multiplied
by a factor $\epsilon ^{1-d}$ that can be absorbed in the definition of
the effective Newton's constant $G_{d+1}$, as in eq. (\ref{Geff}). 
The GB term generates an additional correction to $G_{d+1}$, 
computed in \cite{myers2}. Omitting the factor of 2 that corresponds to the
two copies of AdS in the RS construction, we have
\be
\Gt_{d+1}=(d-1) \epsilon^{d-1}\frac{G_{d+2}  }{(1+2\lx f)\Rt},
\label{Gefft} \ee
which generalizes eq. (\ref{Geff}).

The holographic calculation of the 
entanglement entropy of the dual CFT requires a modification of 
the functional to be extremized, in a way that generalizes the
Wald entropy \cite{wald} for surfaces that do not correspond 
to horizons \cite{camps}. The leading contribution to the 
entanglement entropy is proportional to the area of the 
entangling surface, in units of $\Rt$ \cite{myers3}. In particular, for
an entangling surface with a strip geometry on a flat boundary, the 
leading contribution is given by the
expression \cite{myers3}   
\be
\St_A=\frac{(1+2\lambda f)\Rt^{d}L^{d-1}}{2 (d-1)  \epsilon^{d-1} G_{d+2}},
\label{GBentropy} \ee
where we have translated the result to our notation. It must
be noted that the calculation in \cite{myers3} does not use a RS construction and 
assumes only one copy of AdS space. It is of the same spirit as our 
approach, so that the cutoff can be identified directly with $\epsilon$. 
The subleading contribution to the entropy is similar 
to the second term of eq. (\ref{Rindlerentropy}). It does not 
involve the cutoff $\epsilon$, but depends only on
the width of the strip.
The derivation of the thermal entropy for a Rindler boundary
is completely analogous to the one that lead to eq. (\ref{entropyRindler}).
Dividing the result (\ref{GBentropy}) by 2, for the reasons that we explained
above eq. (\ref{entropyRindler}), and using eq. (\ref{Gefft}),
we obtain
\be
\St_R=\frac{\Rt^{d-1}L^{d-1}}{4\Gt_{d+1}}.
\label{entropyRindlert} \ee
This has the same form as the result (\ref{entropyRindler}),
with the only difference concerning the appearance of the appropriate
physical scales.

The calculation of the de Sitter entropy for a theory dual to a
GB bulk in the limit $\epsilon\to 0$ 
is expected to produce a leading result for the entropy of
the form of eq. (\ref{entropydss}), with the appearance of the
appropriate physical scales.
Possible differences with the Rindler case concern the subleading
corrections. In particular, the presence of subleading terms
with cutoff dependence in the expression for the 
entanglement entropy of a spherical entangling surface 
indicates the appearance of terms suppressed by higher powers of
$G_{d+1}$ in the area law for 
the thermal dS entropy. Such terms become negligible in the
limit $\epsilon \to 0$ or $G_{d+1} \to \infty$. 
Their calculation
goes beyond the scope of this letter and we postpone it for a
future publication.

In conclusion, the holographic method of Ryu and Takayanagi
can reproduce correctly the entropy of spaces with horizons.
The entanglement entropy of regions
that cover the whole space inside the horizons
can be identified with
the standard thermal entropy of these spaces.
For a precise quantitative agreement, one must define
the effective Newton's constant appropriately and account carefully for the way the
AdS space is covered by the coordinates. In this picture the UV divergence of
the entanglement entropy is reflected in the suppression of the effective 
Newton's constant by the infinite volume near the AdS boundary, an IR effect.

\section*{Acknowledgments}
We would like to thank E. Kiritsis, I. Papadimitriou, T. Tomaras and N. Toumbas for 
useful discussions. This work has been funded by 
the Hellenic Foundation for Research and Innovation (HFRI) and 
the General Secretariat for Research and Technology (GSRT)
under grant agreement No 2344.


\end{document}